\def\lap{\hbox{{\lower -2.5pt\hbox{$<$}}\hskip-8pt\raise
-2.5pt\hbox{$\sim$}}}
\def\ga{\hbox{{\lower -2.5pt\hbox{$>$}}\hskip -8pt\raise
-2.5pt\hbox{$\sim$}}}

\documentclass[11pt,dvips]{article}

\usepackage{epsfig,times} 
\usepackage{picinpar}
\usepackage{wrapfig}
\usepackage{floatflt}
\makeatletter
\renewcommand{\section}{\@startsection{section}{1}{0in}
	{0.4\baselineskip}{0.1\baselineskip}{\Large\bf}}
\renewcommand{\subsection}{\@startsection{subsection}{2}{0in}
	{0.25\baselineskip}{-\baselineskip}{\large\bf}}
\renewcommand{\subsubsection}{\@startsection{subsubsection}{3}{0in}
	{0.1\baselineskip}{-\baselineskip}{\normalsize\bf}}
\makeatother
\begin{document}

%
\makeatletter\newcommand{\ps@icrc}{
\renewcommand{\@oddhead}{\slshape{OG.3.3.03}\hfil}}
\makeatother\thispagestyle{icrc}
%


\begin{center}
{\LARGE \bf Galactic Ultra-High-Energy Cosmic Rays}
\end{center}

\begin{center}
{\bf A.V. Olinto$^{1}$, R.I. Epstein$^{2}$, and P. Blasi$^{1}$}\\
{\it $^{1}$Department of Astronomy and Astrophysics, University of Chicago,
Chicago, IL 60637, USA\\
$^{2}$NIS-2, Los Alamos National Laboratory, Los Alamos, NM 87545, USA}
\end{center}

\begin{center}
{\large \bf Abstract\\}
\end{center}
\vspace{-0.5ex}
The absence of the expected GZK cutoff strongly challenges 
the notion that the highest-energy cosmic rays are
of distant extragalactic origin. We discuss the possibility 
that these ultra-high-energy events originate in our Galaxy
and propose that they may be due to iron nuclei accelerated 
from young, strongly magnetic neutron stars. Newly formed pulsars
accelerate ions from their surface through relativistic MHD
winds. We find that pulsars whose initial spin periods are
shorter than $\sim 4 (B_S/10^{13}{\rm G})$ ms, where $B_S$ is  the
surface magnetic field, can accelerate iron ions to greater than $10^{20}
eV$.  These ions can pass through the remnant of the supernova explosion
that produced the pulsar without suffering significant
spallation reactions. Depending on the structure of the 
galactic magnetic field, the trajectories of the
iron ions from galactic sources can be consistent with the
observed arrival directions of the highest energy events.
%

\vspace{1ex}

%
%
\section{Introduction:}
\label{intro.sec}
 
The detection  by AGASA (Hayashida 1994; Takeda et al. 1998, 1999), Fly's
Eye (Bird et al. 1995, 1994, 1993), and Haverah Park (Lawrence, Reid   
\& Watson 1971) of cosmic ray events with energies above the
expected Greisen-Zatsepin-Kuzmin (GZK) cutoff (Greisen 1966; Zatsepin
\& Kuzmin  1966) has triggered considerable interest (see e.g, Cronin
1999; Bhattacharjee \& Sigl 1999 for recent reviews). The  cutoff should be
present if the  ultra-high energy particles are protons from  extragalactic
sources. Cosmic ray protons of energies above
$5
\times 10^{19}$ eV lose their energy to photopion production off the cosmic
microwave background and cannot originate further than about $50\,$Mpc
away from us. Alternatively, if ultra-high-energy cosmic rays (UHECRs)
are protons from sources closer than $50\,$Mpc, the arrival directions of
the events should point toward their source. The present data shows a
mostly  isotropic distribution and no sign of the local distribution of
galaxies or of the galactic disk (Takeda et al. 1998).   In sum, the
origin of these ultra-high energy particles  remains a mystery. 
Here we discuss how the early evolution of young pulsars may be
responsible for the yet unexplained flux of cosmic rays beyond the GZK
cutoff (Blasi, Epstein, \& Olinto 1999).

\section{UHECRs from Newborn Pulsars:}

When neutron stars are born during the collapse of supermassive stars,
they begin their life with very fast rotation ($\Omega \sim 3000 {\rm
s}^{-1}$) and very large magnetic fields (on the star's surface $B_S \ga
10^{13}$ G). Inside the light cylinder (i.e, $r \lap R_{lc}$ where 
$R_{lc} = c /\Omega$) a magnetosphere corotates with  a dipole magnetic
field component that scales as 
$B(r) = B_S (R_S / r)^3$ where the radius of the
star $R_S \simeq 10^6$ cm. The magnetosphere has a density 
(Goldreich \& Julian) 
$n_{GJ}(r) \simeq {B(r) \Omega/4 \pi Z e c}  \ ,
$where  $Z$ is the charge of the
magnetospheric nucleus, $e$ the electric charge and $c$ the speed of light.
As the distance from the star increases, the
dipole field structure cannot be maintained and beyond the {\it
light cylinder} the field is mostly azimuthal.  For young rapidly
rotating neutron stars, the light cylinder is just about ten times the
star radius,  $R_{lc} = 10^7
\Omega_{3k}^{-1} $ cm (where $\Omega_{3k} \equiv \Omega / 3000 \, {\rm
s}^{-1}$).

Iron nuclei stripped by strong electric fields from the surface of the
neutron star,  fill much of the magnetosphere.   From the light
cylinder, a relativistic plasma expands as a magnetohydrodynamic (MHD)
wind (Gallant \& Arons 1994). The field strength 
($B \propto r^{-1}$) and geometry in this region
are such that the  plasma moves relativistically with Alfv\'en speed
close to the speed of light. 
 In the rest frame of the wind, the plasma is
relatively cold while in the star's rest frame the plasma moves with
$\gamma \sim 10^9 - 10^{10}$. 

The typical energy of the accelerated cosmic rays can be estimated by
considering the  magnetic energy per ion at the light cylinder $ E_{cr} =
B_{lc}^2/ 8 \pi n_{GJ}$. At the light cylinder the magnetic
field strength is
$B_{lc} =10^{10} \, {\rm G} \,  B_{13} \Omega_{3k}^3$ and
$ n_{GJ} = 1.7 \times 10^{11} \, {\rm cm}^{-3} \, {B_{13}  \Omega_{3k}^4
/ Z} $  which  gives 
\begin{equation}
 E_{cr} =  4 \times 10^{20}\, {\rm eV}\,  Z_{26} B_{13}
\Omega_{3k}^2 \ , 
\label{eq:Ecr}
\end{equation}
where $ Z \equiv 26 Z_{26}$ and
$B_S \equiv 10^{13}\, {\rm G} \, B_{13}$.
The spectrum of accelerated UHECRs is determined by the evolution of the
 rotational frequency:
As the neutron star
ages, the rotation speed decreases due to electromagnetic and
gravitational radiation. For initial periods of $\ga  1 $ ms  and $B_S
\ga 10^{13}$ G, the spin down is dominated by magnetic dipole radiation 
given by:
$ I \Omega \dot \Omega =  - {B_S^2  R_S^6 \Omega^4 / 6 c^3}$. 

For a moment of inertia $I = 10^{45}$ g cm$^2$, the star's spin down time
is
$t_{sd}\equiv \Omega/ \dot \Omega  \sim  1.8 \times 10^8 {\rm
s} \, B_{13}^{-2} \Omega_{3k}^{-2}$, and the spectrum can be found to be
(Blasi, Epstein, \& Olinto 1999):
\begin{equation}
N(E) dE =  \frac{\dot Q}{\dot \Omega}\, \frac{d\Omega}{dE}\,  dE = {\xi
\, 5.5 \cdot 10^{31} \over B_{13}  E_{20}  Z_{26} }  {\rm GeV}^{-1}   \ .
\label{eq:spec}
\end{equation}
The particle flux is thus 
$ \dot Q = \xi \, n_{GJ} \, \pi R_{lc}^2 c = \xi   {B_{13} \Omega_{3k}^2
\over Z_{26}}  6 \times 10^{34} \,  {\rm s}^{-1}$
where $\xi<1$  is the efficiency  for accelerating 
 particles at the light cylinder.
The energy as a function of spin frequency in Eq.(\ref{eq:Ecr}) gives
${dE}/{d\Omega}= {1.7 \times 10^{-3} } \ {E / \Omega_{3k}}$.

Young neutron stars are most likely distributed in the Galactic disk at
typical distances  $\sim 8$ kpc. If they are produced at a rate
$1/\tau$ where $\tau = 100 ~\tau_2$ yr, the flux of particles on
Earth is
\begin{equation}
F(E)={N(E) \over \pi R^2 \tau} = { \xi  \,  10^{-23}  \over
\tau_2  B_{13} E_{20}  Z_{26} } \, {\rm GeV}^{-1}  {\rm cm}^{-2} {\rm
s}^{-1} \  \ .
\label{eq:spec2} 
\end{equation}
  Assuming an isotropic
distribution of the arrival directions, the AGASA experiment finds a flux
at $10^{20} {\rm eV}$ of  $F(10^{20} {\rm eV}) = 4 \times 10^{-30}~{\rm
GeV}^{-1}  {\rm cm}^{-2} {\rm s}^{-1}$. Therefore, the efficiency 
need only be $\xi \ga 4 \times 10^{-7}$.  Furthermore, the spectrum
derived above is very flat, $N(E)
\propto E^{-1}$, another property in good agreement with the data.

Even though the young neutron stars are surrounded by the remnants of
the presupernova star, the accelerated particles can easily escape the
supernova remnant without significant degradation for a wide range of
initial magnetic fields and spinning rates. The
supernova event that formed the young neutron star also  ejected the
envelope of the original star, making it possible for cosmic rays to
escape. However, as the envelope expands,  the young neutron star spins
down and may become unable to emit particles of the necessary energy. 
Thus, an additional requirement for this scenario is that the column
density of the envelope becomes transparent before the spinning rate of
the neutron star decreases significantly.

To estimate the evolution of the column density of the envelope, consider
a supernova that imparts
$E_{SN}= 10^{51} E_{51}$ erg on the stellar envelope of mass  
$M_{env}= 10  M_1 {\rm M}_{\odot}$.  The envelope then disperses with
a velocity
$ v_e \simeq \left({2 E_{SN} / M_{env}}\right)^{1/2} = 3 \times 10^8 
\left({E_{51}/ M_1}\right)^{1/2} {\rm cm\ s}^{-1} \ .
$
 The column density of the envelope surrounding the neutron star is
$\Sigma \simeq {M_{env} / 4 \pi R_{eff}^2}
$ where $R_{eff} = R_0 +v_e t$. The initial
value $R_0$ is characteristic of pre-supernova stars and  $R_0\sim
10^{13} {\rm cm}\, R_{13}$. We now have
$$
\Sigma  \simeq   {M_{env}\over 4 \pi \left [ R_0 +v_e t\right]^2}
   = 1.6 \times 10^{14} {\rm g\ cm}^{-2}\,  { M_1^2
E_{51}^{-1}\over t^{2} ( 1 + t_e/t)^2} \ ,
$$ where $t$ is in seconds, and
$  t_e = {R_0 / v_e}   = 3 \times 10^4 \, R_{13} \sqrt{
M_1 /  E_{51}} {\rm s} $.   The condition for iron nuclei to traverse
the supernova envelope without significant losses
is that
$\Sigma <\Sigma_t
\simeq 100 \, $ g cm$^{-2}$.  At late times compared to $t_e$, this
``transparency" condition  gives
$ t > t_{tr} = 1.3\times   10^7 M_1
E_{51}^{-1/2}\, {\rm s}$.  

At the same time, the neutron star loses energy at a rate
given by dipole radiation whose solution is
$$
\Omega_{3k}^{2}(t) = {\Omega_{i3k}^{2} \over  [ 1 + t_8 B_{13}^2
\Omega_{i3k}^2  ]}\ ,
$$
where $\Omega_{i3k}$ is the initial spin period and  $t_8 = t / 10^8 {\rm
s}$.
The evolution of the maximum energy is thus given by
$$ E_{cr}(t) =   4 \times 10^{20} {\rm eV}\  { Z_{26} B_{13} \Omega_{i3k}^2
  \over  [ 1 +   t_{8}  B_{13}^2
\Omega_{i3k}^2  ] }.
$$
Since the maximum energy decreases as the source evolves, the
condition that a source could  produce the UHECRs is that $E_{cr}$
exceeds the needed energy when the envelope becomes transparent; i.e.,
$E_{cr}(t_{tr}) > 10^{20} E_{20}$ eV.  This translates into
  the following  condition:
$$
 \Omega_{i} > {3000\, {\rm s^{-1}} \over  B_{13}^{1/2} \left [ 4 Z_{26}
E_{20}^{-1} -  0.13 M_1 B_{13} E_{51}^{-1/2}  \right]^{1/2}}.
$$
>From this equation we obtain upper bounds on the surface magnetic field
strength and the star's initial spin period, $P_i = 2 \pi/\Omega_i$; i.e.,
$  B_{13} <
 31   Z_{26}  E_{51}^{1/2}/M_1 E_{20}$ and $P_i < 8 \pi B_{13}^{1/2}
 Z_{26} E_{20}^{-1}$. For $M_1 = 2 $ and
$E_{20}=E_{51}=Z_{26} = 1$, this is just
$B_{13} < 15.4$ and $P_i \lap 10 $ms, not very restrictive values for a
young neutron star. The allowed regions in the
$B_S$-$\Omega_i$ plane are shown in Figure 1 for $E_{20} = 1$ and 3.
The iron  ejected with energies   $\sim 10^{20}$ eV will reach Earth
after some diffusion through the Galactic and halo magnetic fields. The
gyroradius of these UHECRs  in the Galactic field of strength $B_{gal}$ is
$$  r_B =  {E_{cr}\over Z e B} = 1.4 {\rm \ kpc} Z_{26}^{-1} \left({B_{gal}
 \over 3 \mu {\rm G}}\right) ^{-1} E_{20}  $$  which is a few times the
typical distance to a young neutron star ($\sim $ 8 kpc).  Therefore,
the ultra-high energy iron arriving at the Earth would not point  at the
source. According to Zirakashvili et al. (1998), a Galactic iron source
is consistent with the  arrival direction distribution observed by
AGASA for UHECRs.  
\begin{figwindow}[1,r,%
{\mbox{\epsfig{file=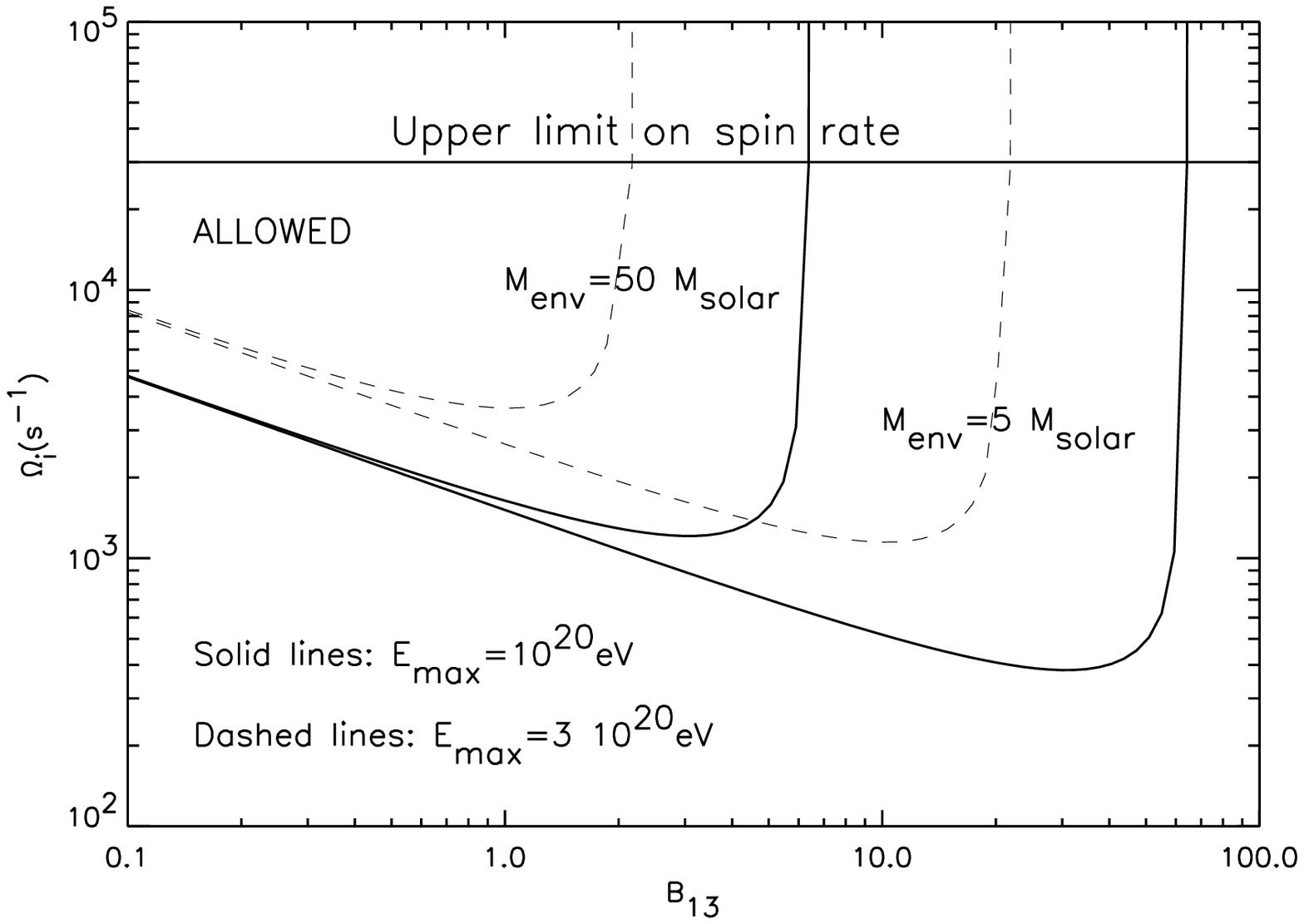,width=4.2in}}},%
{Allowed region of parameters for efficient acceleration.}]
In support of this interpretation, we note that the
cosmic ray component at
$10^{18}$ eV is nearly isotropic. If these cosmic rays are protons of
galactic origin, the
isotropic distribution observed at these energies may be indicative of the
diffusive effect of the Galactic and halo magnetic fields. The iron arrival
distribution  at $10^{20}$ eV probes  similar trajectories to protons at a
few times $10^{18}$ eV. 
Depending on the exact strength and structure of the magnetic field in the
Galaxy and the halo of the Galaxy,  the  ultra-high energy iron may retain
some memory of the source direction.   As the   data improves at the
highest energies a preference for the galactic plane should become  
evident. In addition, the relativistic wind may also accelerate some 
lighter nuclei that can help constrain the scenario and the structure of
the Galactic and halo magnetic fields.   
\end{figwindow}

\section{Conclusion}

We have discussed the possibility that the ultra-high-energy events 
observed past the GZK cutoff  originate in our Galaxy and are due to iron
nuclei accelerated  from young, strongly magnetic neutron stars.
Iron  from the surface of newborn neutron stars are accelerated to
ultra-high energies by a relativistic MHD wind. Pulsars
whose initial spin periods are shorter than $\sim 4 (B_S/10^{13}{\rm G})$
ms can accelerate iron ions to greater than $10^{20} eV$.  These ions can
pass through the remnant of the supernova explosion that produced the
pulsar without suffering significant spallation reactions.

This proposal can be tested by future experiments such as the Auger Project.
The best test of this proposal is a unambiguous composition determination and
a  correlation of arrival directions  for events with energies above
$10^{20}$ eV. Both aspects will be testable with future experiments. Finally,
a detailed study of the lighter element components expected to reach us
from this process will also help constrain this scenario.

{\bf Acknowledgments}

The research was partly supported by NSF through grant AST 94-20759  and
DOE grant DE-FG0291  ER40606 at the  University of Chicago and partly
carried out  under the auspices of the U.S. Department of Energy
with support by IGPP at LANL.

%
%
%
%
\vspace{1ex}
\begin{center}
{\Large\bf References}
\end{center}
%
Bird, D.J. et al., Phys.~Rev.~Lett. 1993, 71  
3401; ApJ  1994, 424, 491; ApJ 1995, 441, 144\\
Bhattacharjee, P., \& Sigl, G. 1999, Phys. Reps. submitted,
astro-ph/9811011\\
Blasi, P., Epstein, R.I., \& Olinto, A.V., 1999, to be submitted\\
Cronin, J.W., Rev. Mod. Phys. 1999, 71, S165 \\ 
Gallant, Y.A., \& Arons, J. 1994, ApJ 435, 230 \\
Greisen, K., Phys. Rev. Lett. 1966, 16, 748 \\
Hayashida, N., Phys. Rev. Lett.  1994,  73, 3491\\
Lawrence, M.A., Reid, R.J.O.
\& Watson, A.A., J.~Phys.~G Nucl.~Part.~Phys. 1991, 17, 733\\
Takeda, M.  et al., Phys.~Rev.~Lett. 1998, 81, 1163; ibid, 1999,  
astro-ph/9902239, submitted to Astrophys.~J.\\
Zatsepin, G.T., \& Kuzmin,V.A, Sov. Phys.-JETP Lett. 1966, 4, 78 \\
Zirakashvili, V.N., Pochepkin, D.N., Ptuskin, V.S., \& Rogovaya,
S.I., Astron. Lett. 1998, 24, 139\\

\end{document}